# Rotation Symmetry Spontaneous Breaking of Edge States in Zigzag Carbon Nanotubes

*Weiliang Wang, Ningsheng Xu, Zhibing Li*[*]

State Key Laboratory of Optelectronic Materials and Technologies, School of Physics and Engineering, Sun Yat-Sen University, Guangzhou 510275, China

## Abstract

Analytical solutions of the edge states were obtained for the ($N$, 0) type carbon nanotubes with distorted ending bonds. It was found that the edge states are mixed via the distortion. The total energies for $N=5$ and $N \geq 7$ are lower in the asymmetric configurations of ending bonds than those having axial rotation symmetry. Thereby the symmetry is breaking spontaneously. The results imply that the symmetry of electronic states at the apex depends on the occupation; the electron density pattern at the apex could change dramatically and could be controlled by applying an external field.

## I. INTRODUCTION

Generally, the performance of nano-devices depends strongly on boundary conditions. The nanotube-based field emission (FE) devices, such as flat panel displays, molecular sensors, and scanning tunneling microscope [1] are typical examples, of which the ending structure plays a decisive role. The process of FE is basically quantum tunneling of electrons through the potential barrier between the nanotube apex and the vacuum. Therefore it is very sensitive to both the applied field and the electronic structure of the apex.[2,3] The present paper is interested in the symmetry of the electron structure at the apex.

---

[*] Corresponding author: stslzb@mail.sysu.edu.cn

The zigzag single-walled carbon nanotube (ZSWCNT) is a graphene with zigzag edge rolled up into a cylinder. It had been known that the ($N$, 0) type ZSWCNT is semimetallic if $N$ has a factor 3, otherwise semiconducting. However, in the FE, the edge states are more relevant.[4] The peculiar edge states (ES) of graphene ribbons with zigzag edges [5,6] have attracted great interest as it had been predicted theoretically that the edge states have surprising electronic and magnetic properties,[7] and as it had been possible to investigate the single-layered graphene experimentally.[8] It is desirable to reinvestigate the edge states of ZSWCNTs on this background since the FE is sensitive to the edge properties.[9-11] Many efforts had been dedicated to the calculation of electronic structure of the SWCNT apex via, for instances, tight binding (TB) method,[3,12] semiclassical modeling,[13] and *ab initio* simulation.[10,14] However, in these studies, the coupling between the ES and the structure distortion has not been fully addressed. The present paper shows that the coupling will break the axial rotation symmetry and has strong effect on the FE image of the ZSWCNT.

The physical origin of the symmetry breaking is resembled to the Peierls instability, or more generally to the Jahn-Teller effect, where no external force is directly involved in the symmetry breaking. In the Peierls instability, the coupling of phonon modes and electron states near Fermi level leads to a lattice distortion that opens a gap between the conduction band and the valence band. In principle this can happen, as had been discussed for the graphene ribbons with zigzag edges.[6,15] However, it had been known that the increase of gap due to the Peierls instability in SWCNTs is greatly suppressed and thus hard to be observed.[12] Instead, we found that the symmetry breaking of the ES is large due to the distortion of carbon-hydrogen bonds (C-H bonds) which terminate the ZSWCNT. The symmetry breaking can be manifested as an asymmetric FE image which would be detectable in the FE experiment. The spontaneous breaking of symmetry as one of the most profound physical concepts has shown its importance in the fields of condensed matters and elementary particles. It should be interesting to see an instance of spontaneous symmetry breaking in nano-systems.

In Section II, the solution of the ES first given by Klein and Fujita *et al.* [6,7] will be recalled and applied to the ZSWCNT, with presumed axial symmetry. In Section III, the interaction potential originated from the distortion of the ending C-H bonds is discussed. As a consequence, the axial rotation symmetry of the ES is broken spontaneously. The last section is devoted to discussions and summary.

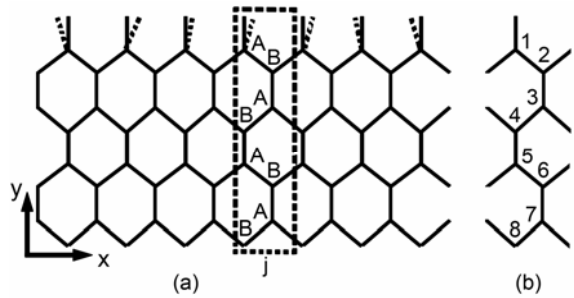

FIG. 1. The (7, 0) ZSWCNT with 8 rows of carbons is presented. The nanotube is spread on a plane. The nodes are carbon atoms, with the top bonds saturated by hydrogen atoms. The dashed lines are the distorted C-H bonds. The j-th unit cell is indicated by the dotted square box in (a), and is re-given in (b) where carbons are labeled by the row number.

## II. EDGE STATES OF ZSWCNT

The atomic structure of the ZSWCNT is presented in Fig.1. The carbon sheet forms a bipartite lattice where the carbons can be divided into A and B sublattices, each of them forms a triangle lattice. The top bonds are saturated by hydrogen atoms. Each row consists of $N$ atoms. The carbons on the edge of the left hand side should be understood as the same as those on edge of the right hand side. Ignoring the curvature effects, the ZSWCNT is treated as the zigzag graphene ribbon with periodic boundary condition, and the Cartesian coordinate as presented in Fig.1 is used. The axial rotation becomes the transversal translation. If $N$ is infinite, this treatment is exact. For finite $N$, as far as local properties are concerned, the ignorance of curvature effects should be acceptable.

The eigenstates of the Hückel model of the zigzag graphene ribbon have been solved and the ES at the Fermi level have been found. [5,6] Their results can be applied to the ZSWCNT directly. However, the continuous transversal lattice wave number of the graphene is replaced by a discrete set of numbers in the ZSWCNT. The main results will be reviewed with our notations in this section.

Only the $2p_z$ states ($z$ direction perpendicular to the plane) will be considered. For a ZSWCNT of $M$ rows of carbons, there are $M$ relevant atomic states, $\{|m_j\rangle\}$ in the j-th cell. Their wave functions are

$$\langle \vec{r}|m_j\rangle = \psi(\vec{r} - \vec{r}_{mj}) \qquad (1)$$

Where $\vec{r}_{mj}$ is the position of the $m$-th carbon in the j-th unit cell. As shown in Fig1 (b), $m$ is also the row label. The wave function $\psi(\vec{r})$ is presumably the $2p_z$ orbital in an effective symmetric crystal potential, with the atomic energy $\varepsilon_0$ (it is set to zero here). For a ($N$, 0) ZSWCNT, $j = 1,2,\cdots,N$. Due to the axial symmetry, the molecular orbitals can be expanded via Bloch waves. The Bloch waves are

$$\Phi_{ml}(\vec{r}) = \frac{1}{\sqrt{N}} \sum_{j=1}^{N} e^{ik_l x_{mj}} \psi(\vec{r} - \vec{r}_{mj}), \qquad (2)$$

The periodic boundary condition in $x$ direction requires that $k_l = 2\pi l /(Na)$, with $l = 0,1,\ldots,N$-1. The sublattice spacing is known as $a$=2.494Å. The molecular orbitals are linear combinations of the Bloch waves

$$\Psi_l(\vec{r}) = \sum_{m=1}^{M} C_{ml}\Phi_{ml}(\vec{r}), \tag{3}$$

The superposition coefficients $\{C_{ml}\}$ are determined by the variational method, $\delta\langle\hat{H}_0\rangle_l = 0$, where $\hat{H}_0$ is the symmetric Hamiltonian,

$$E_l^{(0)} = \langle\hat{H}_0\rangle_l = \frac{\langle\Psi_l|\hat{H}_0|\Psi_l\rangle}{\langle\Psi_l|\Psi_l\rangle} \tag{4}$$

Neglecting the overlap integral, i.e., $\langle\Phi_{m'l}|\Phi_{ml}\rangle = \delta_{m'm}$, the least of $E_l^{(0)}$ leads to a secular problem of a $M \times M$ matrix

$$H_0^{(l)} = \begin{pmatrix} 0 & h_1 & 0 & 0 & 0 & \\ h_1 & 0 & h_2 & 0 & 0 & \\ 0 & h_2 & 0 & h_1 & 0 & \cdots \\ 0 & 0 & h_1 & 0 & h_2 & \\ 0 & 0 & 0 & h_2 & 0 & \\ & \vdots & & & & \ddots \end{pmatrix} \tag{5}$$

For our model, $h_1 = 2t\cos(\varphi_l)$ and $h_2 = t$, with $\varphi_l = k_l a/2$ and $t = \langle m_j|(m+1)_j\rangle$. The eigenvalues of $H_0^{(l)}$ are the energies of the molecular orbitals. In the TB approximation, the transfer integral $t$ is usually determined by fitting the *ab initio* results, which has the value $t = -3.033$ eV for SWCNTs. Note that $k_{N/2} = \pi/a$ is the decoupling point (D point) where $h_1 = 0$. At this point,[7] the electron is exactly localized at the end row of carbons, with the $C_{m,N/2} = -i\delta_{m,1}$ or $C_{m,N/2} = e^{-iW\pi/2}\delta_{m,M}$. Where W=M/2 for even M. This edge state exists only for even N. For the edge states with $k_l$ not at the D point, the eigenvalues are,[6]

$$\begin{aligned} E_l^{(0)\pm} &= \pm\sqrt{h_1^2 + h_2^2 + 2h_1 h_2 \cos\theta_l} \\ &= \mp t\sqrt{1 + 4\cos^2\varphi_l + 4\cos\varphi_l \cos\theta_l} \end{aligned} \tag{6}$$

Where $\theta_l$ satisfies

$$\sin((W+1)\theta_l) + c_l \sin(W\theta_l) = 0 \tag{7}$$

We have defined $c_l = h_2/h_1 = 1/(2\cos\varphi_l)$ and assumed that $M$ is even. There are $M$ real $\theta_l$ being solutions of (7) for $|c_l| < (W+1)/W$. It has been noted by Klein that, as $|c_l|$ increases, the largest solution of $\theta_l$ drifts towards $\pi$ and will move off the real axis into the complex plane, become a complex solution $\theta_l = \pi + i\alpha_l$ when $|c_l| > (W+1)/W$. Correspondingly, the states are localized at either edges, and referred to as edge states (ES). For infinite $W$, $\alpha_l$ is asymptotically given by $\alpha_l = \ln|c_l|$. The criterion for the ES turns out to be $2\pi/(3a) < k_l < 4\pi/(3a)$, i.e., $N/3 < l < 2N/3$, with the boundary values so-called K points corresponding to two degenerated extended states. The energies of the ES approach zero (that is the Fermi level) asymptotically for large $M$. There are two ES for each $k_l$ that satisfies the criterion, corresponding to two edges of the ZSWCNT respectively. For the FE process, only the ES associated to the up-edge is interesting. The normalized amplitudes, for $k_l$ not at the D point, are known as

$$C_{2p-1,l} = c_l^{-p}\sqrt{c_l^2 - 1}(-e^{ik_l a/2})^p, \qquad p = 1, 2, \cdots \qquad (8)$$

$$C_{2p,l} = 0 \qquad , \qquad p = 1, 2, \cdots \qquad (9)$$

These states have the axial symmetry and are localized at the up-edge, hereafter it will be referred to as symmetric up-edge states (SUES). At zero temperature, each SUES is occupied by one electron (either spin up or spin down).

Denote the number of independent edge states as $d_e$, that is the dimension of the edge state space. In the range of 2<N<11, $d_e = 3$ for $N$=8 and 10; $d_e = 2$ for $N$=5, 7, and 9; $d_e = 2$ for $N$=4 and 6; $d_e = 0$ for $N$=3.

III. SPONTANEOUS SYMMETRY BREAKING

Klein has argued that the Peierls gap does not exist for infinite wide graphene ribbon (here, it is corresponding to an infinite long ZSWCNT). [6] However, since the SUES of different $k_l$ are almost degenerate, they will mix under the distortion of the ending structure and lead to a gap via the Jahn-Teller mechanism. Now let us assume that the hydrogen atoms have small displacements with respect to their symmetric positions (the H-ring distortion), as indicated in Fig.1 (a) by the dashed lines. Since the C-H bonds are polarized and charged, [16] the electrons in the molecular orbitals will feel an extra potential proportional to the displacements of the hydrogen atoms (e.g., through the dipoles and polarization of the C-H bonds), besides the symmetric crystal potential.

3.1 The distortion potential

Let the displacement of the j-th hydrogen be $\vec{q}_j$. The H-ring as a 3-dimensional object has three vibration modes. The longitudinal mode (in *y* direction) causes more elastic energy, so it will be ignored. On the perpendicular plane, there are radial mode and twisting mode, as depictured in Fig. 2 (a), and (b), respectively. The distortion potential acting on the π-orbitals can be written as

$$V(\vec{r}) = -\lambda \sum_{j=1}^{N} \vec{q}_j \cdot (\vec{r} - \vec{r}_{0j}) f(\vec{r} - \vec{r}_{0j}) \qquad (10)$$

Where $\vec{r}_{0j}$ is the coordinate of the j-th hydrogen atom. The function $f(\vec{r}) = f(x, y, z)$ is even in the *x* and *z* reflection, and it decreases rapidly as $|\vec{r}|$ increases. Since we neglected the longitudinal mode, $\vec{q}_j$ has only the *x*- and *z*-components $\{q_j^{x,z}\}$. The potential *V* describes the coupling between the lattice variables $\{\vec{q}_j\}$ and the π-electronic variable ($\vec{r}$). Note that *V* is invariant under the $C_N$ axial rotation (including lattice and electronic variables, i.e, the combinational transformation: $x \to x+a$ and $q_j^{x,z} \to q_{j-1}^{x,z}$).

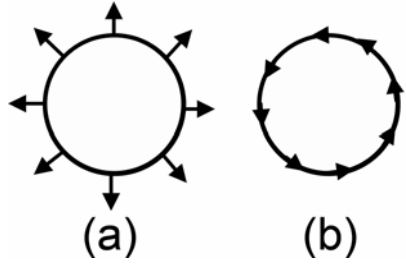

FIG. 2. The distortion modes of the hydrogen ring. (a) The radial breathing mode. (b) The twisting mode. The tube axis is perpendicular to the circles.

The displacement of the j-th hydrogen atom can be expanded as

$$\vec{q}_j = \frac{1}{2}\sum_{l=0}^{N-1}\left(\vec{Q}_l e^{ik_l x_j} + \vec{Q}_l^* e^{-ik_l x_j}\right) \qquad (11)$$

As a consequence of this distortion, the SUES will mix with each other subject to the symmetry. Because $V$ is a short range interaction, it only affects atomic states that are near to the hydrogen (Fig.3). The extended states has very small amplitude at the edge sites, so will not be affected by the distortion of the ending bonds.

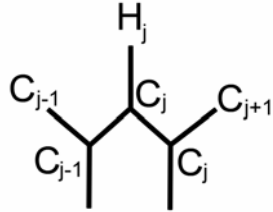

FIG. 3. The $C_j$ is the carbons nearest to the hydrogen $H_j$, $C_{j\pm1}$ are the second nearest to $H_j$, and $C'_j$ and $C'_{j-1}$ are the third nearest to $H_j$.

We first consider the x-component of the H-ring distortion. It is useful to remember that the wave function $\psi(\vec{r}-\vec{r}_{1(j+1)})$ transforms to $\psi(\vec{r}-\vec{r}_{1(j-1)})$ in the x-reflection transformation with respect to $x_{1j}$. Due to this fact, the uniform distortion in x direction does not affect the SUES in the first order approximation. The relevant part of the potential of x-component distortion that has non-zero matrix elements in the SUES space is

$$V_{rel}^x = -\frac{\lambda}{2} \sum_{j=1}^{N} \sum_{l=1}^{d_e-1} \left( Q_l^x e^{ik_l x_j} + Q_l^{x*} e^{-ik_l x_j} \right)(x - x_{0j}) f(\vec{r} - \vec{r}_{0j}) \qquad (12)$$

Before the H-ring distortion emerges, it is nature to suppose that $x_{0j}$ of the j-th hydrogen atom are the same as the j-th carbon in the first row, i.e., $x_{0j} = x_{1j}$. Define

$$u = \left\langle \psi(\vec{r} - \vec{r}_{1j}) \middle| (x - x_{1j}) f(\vec{r} - \vec{r}_{0j}) \middle| \psi(\vec{r} - \vec{r}_{1j+1}) \right\rangle \qquad (13)$$

Due to the *x*-reflection symmetry, one has

$$\left\langle \psi(\vec{r} - \vec{r}_{1j}) \middle| (x - x_{1j}) f(\vec{r} - \vec{r}_{0j}) \middle| \psi(\vec{r} - \vec{r}_{1j-1}) \right\rangle = -u \qquad (14)$$

The matrix elements of $V_{rel}^x$ between two SUES are

$$\Omega_{\bar{l}l} \equiv \left\langle \Psi_{\bar{l}} \middle| V_{rel}^x \middle| \Psi_l \right\rangle$$
$$= -i2\lambda u C_{1\bar{l}}^* C_{1l} \sin \frac{k_{l-\bar{l}} a}{2} \cos \frac{k_{\bar{l}+l} a}{2} \sum_{l'=1}^{d_e-1} \left( Q_{l'}^x \delta_{\bar{l},l+l'} + Q_{l'}^{x*} \delta_{\bar{l},l-l'} \right) \qquad (15)$$

The delta-functions are originated from the rotation symmetry. For $l > \bar{l}$, using (8), one has

$$\Omega_{\bar{l}l} = -i2\lambda u f_{\bar{l}l} e^{ik_{l-\bar{l}} a/2} Q_{l-\bar{l}}^{x*} \sin \frac{k_{l-\bar{l}} a}{2} \cos \frac{k_{\bar{l}+l} a}{2} \qquad (16)$$

When $l < \bar{l}$, $Q_{l-\bar{l}}^{x*}$ is replaced by $Q_{\bar{l}-l}^x$. We have defined $f_{\bar{l}l} = \sqrt{(c_{\bar{l}}^2 - 1)(c_l^2 - 1)}/(c_{\bar{l}} c_l)$ if $\bar{l}, l \neq N/2$ and at the D point, $f_{\bar{l},N/2} = \sqrt{(c_{\bar{l}}^2 - 1)}/c_{\bar{l}}$. The diagonal element $\Omega_{ll} = 0$.

Now, we consider the *z*-component distortion. It is expected that the C-H bonds are polarized and charged. Due to the repulsive interaction of the ending bonds, the C-H bonds will bend outward, i.e., leading to a uniform radial deformation, $\delta = z_{0j} - z_{1j} > 0$. The uniform radial deformation does not break the axial symmetry and its effect has been accounted in the symmetric atomic orbitals. Therefore, the *z*-component distortion emerged on the background of a fixed uniform radial deformation. In the other words, the relevant part of the potential of the *z*-component distortion has no contribution from the uniform radial deformation of the H-ring. Its general form is

$$V_{rel}^z = -\frac{\lambda}{2} \sum_{j=1}^{N} \sum_{l=1}^{d_e-1} \left( Q_l^z e^{ik_l x_j} + Q_l^{z*} e^{-ik_l x_j} \right)(z - z_{0j}) f(\vec{r} - \vec{r}_{0j}) \quad (17)$$

The matrix element between two SUES states $|\Psi_{\bar{l}}\rangle$ ans $|\Psi_l\rangle$, with $l > \bar{l}$, reads

$$\Sigma_{\bar{l}l} \equiv \langle \Psi_{\bar{l}} | V_{rel}^z | \Psi_l \rangle = -\frac{\lambda}{2} C_{1\bar{l}}^* C_{1l} Q_{l-\bar{l}}^{z*} \frac{1}{N} \sum_{j,j',j''=1}^{N} v_{j'jj''} e^{-ik_{\bar{l}} x_{1j'} + ik_l x_{1j''} - ik_{l-\bar{l}} x_{1j}} \quad (18)$$

We have defined

$$v_{j'jj''} = \langle \psi(\vec{r} - \vec{r}_{1j'}) | (z - z_{0j}) f(\vec{r} - \vec{r}_{0j}) | \psi(\vec{r} - \vec{r}_{1j''}) \rangle \quad (19)$$

As a consequence of the nonzero $\delta$, the matrix element $v_{jjj}$ is nonzero. The off-diagonal elements with $j' \neq j$ or $j'' \neq j$ will be neglected. Since $\delta$ is small, one can expand $v_{jjj}$ in $\delta$ and keep the leading term. Due to the symmetry, the leading term is proportional to $\delta$, i.e., $v_{jjj} = v\delta$, with

$$v = -\langle \psi(\vec{r} - \vec{r}_{1j}) | f(x - x_{1j}, y - y_{0j}, z - z_{1j}) | \psi(\vec{r} - \vec{r}_{1j}) \rangle + \langle \psi(\vec{r} - \vec{r}_{1j}) | (z - z_{1j}) \frac{\partial f(x - x_{1j}, y - y_{0j}, z - z_{1j})}{\partial z_{1j}} | \psi(\vec{r} - \vec{r}_{1j}) \rangle \quad (20)$$

Then (18) becomes

$$\Sigma_{\bar{l}l} = -\frac{1}{2} \lambda v \delta f_{\bar{l}l} Q_{l-\bar{l}}^{z*} e^{ik_{l-\bar{l}} a/2} \quad (21)$$

When $l < \bar{l}$, $Q_{l-\bar{l}}^{z*}$ is replaced by $Q_{\bar{l}-l}^z$. The diagonal elements are zero.

Following the standard perturbation theory for degenerate states, the physical realized edge states can be found by diagonalizing the matrix $V_{rel} = V_{rel}^x + V_{rel}^z$, which has elements $V_{\bar{l}l} = \Omega_{\bar{l}l} + \Sigma_{\bar{l}l}$, with $N/3 < \bar{l}(l) < 2N/3$ and $|\bar{l} - l| < d_e$.

For the edge state space of $d_e = 1$, i.e., for N=4 and 6, there is only one edge state and no symmetry breaking. We will focus on the ZSWCNT of two-dimensional and three-dimensional edge state space. The former includes N=5,7, and 9. The latter includes N=8,10, and 12. The distortion for larger N is small as the displacement has a factor 1/N (section 4).

3.2 Two-dimensional edge state space

The edge state space has the bases $\{\Psi_{l_0}, \Psi_{l_0+1}\}$. The distortion mixes them to form the physical states

$$\Psi^{\pm} = \frac{1}{\sqrt{2}}\left(\Psi_{l_0} \pm \Psi_{l_0+1}\right) \tag{22}$$

The corresponding energies are $E^{\pm} = \pm |V_{l_0,l_0+1}|$, with

$$|V_{l_0,l_0+1}| = |\lambda|\left(2\cos(\frac{\pi}{N})-1\right)\left|i2uQ_1^{x*}\sin(\frac{\pi}{N}) - \frac{1}{2}v\delta Q_1^{z*}\right| \tag{23}$$

The level $E^- < 0$ is double occupied. Therefore, the electronic energy gain via the H-ring distortion is $\Delta_e = 2|E^-|$.

3.3 Three-dimensional edge state space

The edge state space has the bases $\{\Psi_{N/2-1}, \Psi_{N/2}, \Psi_{N/2+1}\}$, where $\Psi_{N/2}$ is the D point edge state. The matrix elements are

$$\begin{aligned} V_{N/2-1,N/2} &= -V_{N/2,N/2+1} \\ &= \lambda e^{ik_1 a/2}\sqrt{\left(2\cos(\frac{2\pi}{N})-1\right)}\left(iuQ_1^{x*}\sin(\frac{2\pi}{N}) - \frac{1}{2}v\delta Q_1^{z*}\right) \end{aligned} \tag{24}$$

$$V_{N/2-1,N/2+1} = \lambda e^{ik_2 a/2}\left(2\cos(\frac{2\pi}{N})-1\right)\left(iuQ_2^{x*}\sin(\frac{\pi}{N}) - \frac{1}{2}v\delta Q_2^{z*}\right) \tag{25}$$

The deduced distortion matrix is written as

$$V_{rel} = \begin{pmatrix} 0 & Ae^{i\phi_A} & Be^{i\phi_B} \\ Ae^{-i\phi_A} & 0 & -Ae^{i\phi_A} \\ Be^{-i\phi_B} & -Ae^{-i\phi_A} & 0 \end{pmatrix} \tag{26}$$

Where $V_{N/2-1,N/2} \equiv Ae^{i\phi_A}$ and $V_{N/2-1,N/2+1} \equiv Be^{i\phi_B}$. The eigenvalues are the solutions of the equation

$$E^3 - (2A^2 + B^2)E + 2A^2 B\cos(2\phi_A - \phi_B) = 0 \tag{27}$$

Define

$$\Lambda_0 = (2A^2 + B^2)^{1/2} \qquad (28)$$

$$\gamma = anti\cos\left(-\frac{3^{3/2} A^2 B \cos(2\phi_A - \phi_B)}{\Lambda_0^3}\right) \qquad (29)$$

where $0 \leq \gamma \leq \pi$. The solutions of (27) can be written as

$$E_1 = \frac{2}{\sqrt{3}} \Lambda_0 \cos\frac{\gamma}{3} \qquad (30)$$

$$E_2 = -\frac{2}{\sqrt{3}} \Lambda_0 \cos\frac{\gamma - \pi}{3} \qquad (31)$$

$$E_3 = -\frac{2}{\sqrt{3}} \Lambda_0 \cos\frac{\gamma + \pi}{3} \qquad (32)$$

One can see that $E_1 > 0$. If $\gamma > \pi/2$, $E_2 < 0$ and $E_3 > 0$. While $\gamma < \pi/2$, $E_2 < E_3 < 0$. The occupation of these levels depends on the chemical energy, $\mu$, that is the energy for adding one electron to the apex. It can be adjusted by the local electric field. In principle, $\mu$ also includes the correlation energy of electrons, which has been neglected in the present paper.

If both $E_2$ and $E_3$ are double occupied by electrons, the electronic energy gain is $\Delta_e^{(a)} = -2(E_2 + E_3) - \mu$, which increases monotonically as $\gamma$ decreases. For the sake of maximizing $\Delta_e^{(a)}$, $\gamma$ should have the smallest value, therefore, $2\phi_A - \phi_B = \pi$. The eigenvalues have explicit expressions with this phase relation,

$$E_a^0 = -B \qquad (33)$$

$$E_a^\pm = \frac{1}{2}(B \pm \Lambda_1) \qquad (34)$$

where $\Lambda_1 = \sqrt{8A^2 + B^2}$. The corresponding eigenvectors are

$$\Psi_a^0 = \frac{1}{\sqrt{2}}\left(e^{i\phi_A}\Psi_{N/2-1} + e^{-i\phi_A}\Psi_{N/2+1}\right) \qquad (35)$$

$$\Psi_a^\pm = \frac{1}{\Lambda^\pm}\left[(\Lambda_0^2 \pm B\Lambda_1)e^{i\phi_A}\Psi_{N/2-1} + A(3B \pm \Lambda_1)\Psi_{N/2} - (\Lambda_0^2 \pm B\Lambda_1)e^{-i\phi_A}\Psi_{N/2+1}\right] \qquad (36)$$

Where $\Lambda^{\pm} = \sqrt{2\Lambda_1^2(A^2+2B^2) \pm 2B(7A^2+2B^2)\Lambda_1}$ is the normalization factor.

If $E_2$ is double occupied while $E_3$ is single occupied, the electronic energy gain is $\Delta_e^{(b)} = -2E_2 - E_3$, which has a maximum at $\gamma = \pi/2$. This condition can be realized by (i) $2\phi_A - \phi_B = \pi/2$; (ii) $A = 0$; or (iii) $B = 0$. The eigenvalues are

$$E_b^0 = 0 \qquad (37)$$

$$E_b^{\pm} = \pm \Lambda_0 \qquad (38)$$

The corresponding eigenvectors are

$$\Psi_b^0 = \frac{1}{\Lambda_0}\left(Ae^{i\phi_A}\Psi_{N/2-1} + iB\Psi_{N/2} + Ae^{-i\phi_A}\Psi_{N/2+1}\right) \qquad (39)$$

$$\Psi_b^{\pm} = \frac{1}{2\Lambda_0}\left[(iB \pm \Lambda_0)e^{i\phi_A}\Psi_{N/2-1} + 2A\Psi_{N/2} + (iB \mp \Lambda_0)e^{-i\phi_A}\Psi_{N/2+1}\right] \qquad (40)$$

3.4 Elastic energy and stable distortion

Since the distortion causes elastic energy, the components of displacements that make no electronic energy gain have zero values. For small distortion, the elastic energy reads,

$$\begin{aligned}\Delta_{elast} &= \frac{1}{2}K\sum_{j=1}^{N}q_j^2 \\ &= \frac{1}{4}NK\sum_{l=1}^{d_e-1}\left(|Q_l^x|^2 + |Q_l^z|^2\right)\end{aligned} \qquad (41)$$

Where $K$ is the spring constant of bending a C-H bond. The total energy for the distortion is the elastic energy subtracted by the electronic energy gain,

$$\Delta E_t(\vec{Q}) = \Delta_{elast} - \Delta_e. \qquad (42)$$

By minimizing the total energy, one can find the stable distortion.

Note that the elastic energy does not depend on the phases of the distortion components. Therefore one can choose the phases to maximize the electronic energy gain, as we have done for $2\phi_A - \phi_B$ in 3.3. Since the electronic energy gain increases monotonically with $|V_{\tilde{i}l}|$, the phases of $Q^x_{|\bar{l}-l|}$ and $Q^z_{|\bar{l}-l|}$ (denoted by $\phi_x$ and $\phi_z$ respectively) should maximize $|V_{\bar{l}l}|$.

Recalling (16) and (21), the requisite of maximum $|V_{\bar{l}l}|$ leads to $\phi_x - \phi_z = -\pi/2$, so

$$|V_{\bar{l}l}| = a_x |Q^x_{|\bar{l}-l|}| + a_z |Q^z_{|\bar{l}-l|}| \quad (43)$$

with $a^x_{\bar{l}l} = |\lambda u f_{\bar{l}l} \sin\frac{k_{|\bar{l}-l|}a}{2} \cos\frac{k_{\bar{l}+l}a}{2}|$ and $a^z = |\lambda v \delta|$.

For ZSWCNT with $d_e = 2$, the elastic energy reads,

$$\Delta^{(2)}_{elast} = \frac{N}{4} K \left[ |Q^z_1|^2 + |Q^x_1|^2 \right] \quad (44)$$

From $\delta E_t / \delta |Q^{x,z}_l| = 0$, one obtains

$$Q^z_1 = \frac{(n_- - n_+)(2\cos(\pi/N) - 1)}{NK} |\lambda v \delta| e^{i\phi} \quad (45)$$

$$Q^x_1 = \frac{8(n_- - n_+)\sin(\pi/N)\cos(3\pi/N)}{NK} |\lambda u| e^{i(\phi - \pi/2)} \quad (46)$$

Where $\phi$ is an arbitrary phase, $n_\pm = 0, 1, 2$ the number of electrons in the state $\Psi^\pm$ respectively. Note that the symmetry is recovered when both levels are double occupied.

For ZSWCNT with $d_e = 3$, the elastic energy reads,

$$\Delta^{(3)}_{elast} = \frac{NK}{4} \left[ |Q^z_1|^2 + |Q^x_1|^2 + |Q^z_2|^2 + |Q^x_2|^2 \right] \quad (47)$$

It is convenience to use $A$, $B$, $|Q^z_1|$, and $|Q^z_2|$ as independent variables. From (43),

$$A = a^x_{N/2-1,N/2} |Q^x_1| + a^z |Q^z_1| \quad (48)$$

$$B = a^x_{N/2-1,N/2+1} |Q^x_2| + a_z |Q^z_2| \quad (49)$$

Eliminating $|Q^x_1|$ and $|Q^x_2|$ from (47) by (48) and (49), $\Delta^{(3)}_{elast}$ becomes a function of $A$, $B$, $|Q^z_1|$, and $|Q^z_2|$. On the other hand, the electronic energy gain is independent of $|Q^z_1|$ or $|Q^z_2|$, therefore, the requisite of the least total energy leads to

$$|Q_1^z| = \frac{a^z}{(a_{N/2-1,N/2}^x)^2 + (a^z)^2} A \qquad (50)$$

$$|Q_2^z| = \frac{a^z}{(a_{N/2-1,N/2+1}^x)^2 + (a^z)^2} B \qquad (51)$$

Substitute them into (48) and (49), one obtains

$$|Q_1^x| = \frac{a_{N/2-1,N/2+1}^x}{(a_{N/2-1,N/2}^x)^2 + (a^z)^2} A \qquad (52)$$

$$|Q_2^x| = \frac{a_{N/2-1,N/2+1}^x}{(a_{N/2-1,N/2+1}^x)^2 + (a^z)^2} B \qquad (53)$$

Substitute (50)-(53) into (47), the elastic energy is written as

$$\Delta_{elast}^{(3)} = \frac{NK}{4((a_{N/2-1,N/2}^x)^2 + (a^z)^2)} A^2 + \frac{NK}{4((a_{N/2-1,N/2+1}^x)^2 + (a^z)^2)} B^2 \qquad (54)$$

(a) Both $E_2$ and $E_3$ double occupied

The electronic energy gain is $\Delta_e^{(a)}(2\phi_A - \phi_B = -\pi) = B + \sqrt{8A^2 + B^2}$. The minimum of total energy gives

$$\frac{NKA}{2[(a_{N/2-1,N/2}^x)^2 + (a^z)^2]} = \frac{8A}{\sqrt{8A^2 + B^2}} \qquad (55)$$

$$\frac{NKB}{2[(a_{N/2-1,N/2+1}^x)^2 + (a^z)^2]} = 1 + \frac{B}{\sqrt{8A^2 + B^2}} \qquad (56)$$

If $A \neq 0$, combine (55), (56), and (50)-(53), one obtains

$$Q_1^z = \frac{32 a^z e^{i\phi}}{\sqrt{2}NK[8(a_{N/2-1,N/2}^x)^2 - (a_{N/2-1,N/2+1}^x)^2 + 7(a^z)^2]} \\ \cdot \sqrt{[(a_{N/2-1,N/2}^x)^2 + (a_{N/2-1,N/2+1}^x)^2][4(a_{N/2-1,N/2}^x)^2 - (a_{N/2-1,N/2+1}^x)^2 + 3(a^z)^2]} \qquad (57)$$

$$Q_1^x = \frac{32 a_{N/2-1,N/2}^x e^{i(\phi-\pi/2)}}{\sqrt{2}NK[8(a_{N/2-1,N/2}^x)^2 - (a_{N/2-1,N/2+1}^x)^2 + 7(a^z)^2]} \\ \cdot \sqrt{[(a_{N/2-1,N/2}^x)^2 + (a_{N/2-1,N/2+1}^x)^2][4(a_{N/2-1,N/2}^x)^2 - (a_{N/2-1,N/2+1}^x)^2 + 3(a^z)^2]} \qquad (58)$$

$$Q_2^z = \frac{16a^z[(a_{N/2-1,N/2}^x)^2 + (a^z)^2]}{NK[8(a_{N/2-1,N/2}^x)^2 - (a_{N/2-1,N/2+1}^x)^2 + 7(a^z)^2]} e^{i(2\phi+\pi)} \qquad (59)$$

$$Q_2^x = \frac{16a_{N/2-1,N/2+1}^x[(a_{N/2-1,N/2}^x)^2 + (a^z)^2]}{NK[8(a_{N/2-1,N/2}^x)^2 - (a_{N/2-1,N/2+1}^x)^2 + 7(a^z)^2]} e^{i(2\phi+\pi/2)} \qquad (60)$$

The phases are added according to the previous arguments. Another solution of (55) is $A = 0$, which is not a minimum point of the total energy.

(b) State $E_3$ single occupied and $E_2$ double occupied

The electronic energy gain is $\Delta_e^{(b)}(2\phi_A - \phi_B = \pi/2) = 2\sqrt{2A^2 + B^2}$. The minimum of total energy gives

$$\frac{NKA}{2[(a_{N/2-1,N/2}^x)^2 + (a^z)^2]} = \frac{4A}{\sqrt{2A^2 + B^2}} \qquad (61)$$

$$\frac{NKB}{2[(a_{N/2-1,N/2+1}^x)^2 + (a^z)^2]} = \frac{2B}{\sqrt{2A^2 + B^2}} \qquad (62)$$

Generally, Eq.(61)-(62) can not be fulfilled simultaneously, except $A = 0$ or $B = 0$. For $A = 0$ and $B \neq 0$, one has $Q_1^z = Q_1^x = 0$ and

$$B = \frac{4[(a_{N/2-1,N/2+1}^x)^2 + (a^z)^2]}{NK} \qquad (63)$$

It can be verified that this is not a minimum point of total energy. For $B = 0$ and $A \neq 0$, one has $Q_2^z = Q_2^x = 0$ and

$$A = \frac{8[(a_{N/2-1,N/2}^x)^2 + (a^z)^2]}{\sqrt{2}NK} \qquad (64)$$

This is a minimum point of the total energy. Therefore, there is only $l = 1$ distortion in the case that the lowest edge state is double occupied and the second lowest edge state is single occupied. The corresponding distortion displacements are

$$Q_1^z = \frac{8a^z}{\sqrt{2}NK} e^{i\phi} \qquad (65)$$

$$Q_1^x = \frac{8a_{N/2-1,N/2}^x}{\sqrt{2NK}} e^{i(\phi-\pi/2)} \qquad (66)$$

IV. DISCUSSIONS AND SUMMARY

In summary, the theory suggests that the axial rotation symmetry of the ($N$, 0) SWCNTs with its opened end saturated by hydrogen atoms is broken spontaneously if $N=5$ or $N \geq 7$, due to the coupling between the edge states of the electron and the distortion of the H-ring that ending the tube. There are degenerate configurations of the distorted H-ring that minimize the total energy, as reflected by an arbitrary continuous phase ($\phi$). It is found that the phase different between the $x$ and $z$ distortion is fixed to $\pi/2$. The $z$-component distortion is possible only if the radius of the H-ring is different from the radius of the ZSWCNT.

The general distortion matrix has been deduced. The distortion does not break the symmetry for $N<5$ and $N=6$. On the other hand, the distortion is smaller for larger $N$. The energy levels and the wave functions of the edge states for nontrivial ZSWCNTs under the distortion of ending bonds with $N \leq 10$ and $N=12$ have been explicitly given. For $N=5$, 7, and 9, there are two edge states on one end of the tube. The only distortion mode has the wave number $\pi/Na$. When both edge states are double occupied, the axial rotational symmetry will be recovered; otherwise, the symmetry is breaking. For $N=8, 10$, and 12, there are three edge states on one end of the tube. The possible distortion would have wave numbers $\pi/Na$ and $2\pi/Na$ if the lowest two edge states are double occupied. However, when the lowest edge state is double occupied while the second lowest one is single occupied, only $\pi/Na$ is possible.

Since the symmetry is sensitive to the occupation of the edge states, we predict that the symmetry will be changed as the neutral apex is charged and lead to a sudden change of electron density pattern at the apex of the tube. Note that the occupation depends on the chemical energy of adding an electron to the apex, and the chemical energy depends on the local electric field at the edge and ultimately on the applied field. This would explain the change of FE pattern with applied field that had been observed in the experiment. [11]

A rough estimation for the spring constant K is 3.7 eV/ $Å^2$ (Recalling that the vibration mode is about 1000cm$^{-1}$ for trans-polyacetylene). Supposing the distortion is resemble to the dipole potential of a charge $\chi e$ shifted by a distance $q$, then the order of $\lambda u$ should be the same as $\chi e^2/(4\pi\varepsilon_0 a^2) \sim \chi E_h(a_{Bohr}/a^2) \sim 1.7 \chi$ eV/Å, where $E_h$ is the Hartree energy and $a_{Bohr}$ the Bohr radius, while $\chi < 1$ takes into account the effective charge and the overlap of atomic wave functions. Then the x-component of the distortion is in the order of $\lambda u/K \sim 0.46\chi$ Å. If the factor $\chi \sim 0.1$, as the order of the second nearest neighbor transfer enregy divided by the Hartree energy, one expects the distortion amplitudes in the order of 0.05Å.

Although the amplitudes of the distortion would be very small, but it could have profound impact on the symmetry of the local density of states, as the mixing of edge states exists in the limitation of zero amplitude of the distortion. As edge states are most relevant to the field electron emission experiment, we thus propose here that the relation between the observable FE images and the edge states would provide a technique to probe the quantum states of nanostructures, which would be helpful for extracting information on both charge density and phase of the states.

ACKNOWLEDGEMENT

The authors thank Z.G. Shuai and J. Iliopoulos for the valuable discussions. The project is supported by the National Natural Science Foundation of China (Grant Nos. 10674182, 90103028, and 90306016) and National Basic Research Program of China (2007CB935500).